# ELECTROSTATIC ACTUATORS OPERATING IN LIQUID ENVIRONMENT: SUPPRESSION OF PULL-IN INSTABILITY AND DYNAMIC RESPONSE

*Anne-Sophie Rollier, Marc Faucher, Bernard Legrand, Dominique Collard and Lionel Buchaillot*

Institut d'Electronique, de Microélectronique et de Nanotechnologie – IEMN CNRS UMR 8520
ISEN Dpt., Silicon Microsystems Group
Cité Scientifique, Av. H. Poincaré, 59652 Villeneuve d'Ascq, FRANCE

## ABSTRACT

This paper presents results about fabrication and operation of electrostatic actuators in liquids with various permittivities. In the static mode, we provide experimental and theoretical demonstration that the pull-in effect can be shifted beyond one third of the initial gap and even be eliminated when electrostatic actuators are operated in liquids. This should benefit to applications in microfluidics requiring either binary state actuation (e.g. pumps, valves) or continuous displacements over the whole gap (e.g. microtweezers). In dynamic mode, actuators like micro-cantilevers present a great interest for Atomic Force Microscopy (AFM) in liquids. As this application requires a good understanding of the cantilever resonance frequency and Q-factor, an analytical modeling in liquid environment has been established. The theoretically derived curves are validated by experimental results using a nitride encapsulated cantilever with integrated electrostatic actuation. Electrode potential screening and undesirable electrochemistry in dielectric liquids are counteracted using AC-voltages. Both experimental and theoretical results should prove useful in micro-cantilever design for AFM in liquids.

## 1. INTRODUCTION

Air and vacuum operation of electrostatic actuators are well controlled as well as their technological fabrication. Some systems have been commercialized such as the Digital Micromirror Device (DMD) invented in 1987 by Texas instrument which is a fast reflective digital light electrostatic switch. However, one of the most ambitious applications would be to make them usable in a liquid environment. This should enable original biomedical applications such as cellular handling and characterization, DNA manipulation or device motion in liquid environment. Up to now, only a few devices have been actuated in liquid solutions, such as electrostatic comb-drives insulated using native silicon dioxide ($SiO_2$) [1]. This paper proposes a process realization based on low stress silicon nitride ($Si_xN_y$) encapsulation of parallel plate electrostatic actuators. Under specific conditions, the actuation in liquid is stable along the whole gap range, which differs from actuation in air where the instability occurs at one third of initial gap. This phenomenon is very interesting in order to control the actuator displacement in the whole range of the electrostatic gap. Furthermore, cantilevers actuated in dynamic mode enable applications such as frequency variation measurements while scanning an object in liquid solutions. The resonance and the quality factor are the limiting parameters of scanning acquisition. To properly image an object in fluidic environment, Q-factor and frequency have to remain as high as possible. This shows the need for precise modeling of mechanisms leading to resonance frequency shift and Q decrease.

## 2. ANALYTICAL MODELING OF ACTUATION

In this study, the parallel plate electrostatic actuator is assumed to be a rectangular cantilever with a length, $L$, a thickness, $t$ and a width, $w$.

### 2.1. Static study

For the quasi-static case, the cantilever is modeled by a capacitor having a fixed electrode while the second one is connected to a mechanical spring. This spring is free to move vertically once an electrostatic force is applied, see Fig.1.

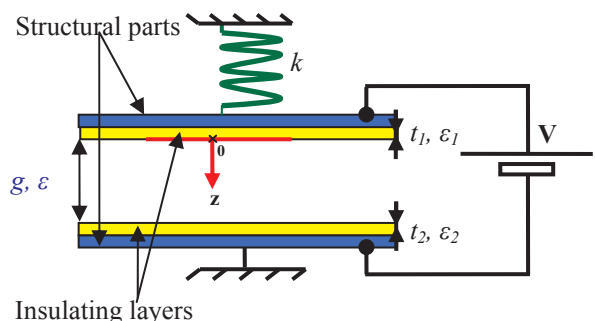

*Fig. 1 Schematic view of parallel electrostatic actuation.*

The electrostatic force acts to reduce the gap between the two electrodes. The restoring mechanical force that is exerted by the spring on the mobile electrode is proportional to the spring stiffness and to its deformation [2]. The equilibrium position of the actuator is reached once the electrostatic and mechanical forces compensate:

$$F_{elect} + F_{mecha} = 0 \Leftrightarrow \frac{1}{2} \times \frac{\varepsilon_0 \varepsilon \times S \times V^2}{\left[ \frac{t_1 \varepsilon}{\varepsilon_1} + \frac{t_2 \varepsilon}{\varepsilon_2} + (g - z) \right]^2} = k \times z \quad (1)$$

Where, $k$ is the spring constant, $g$ the gap between the electrodes, $V$ the supply voltage, $z$ the displacement of the mobile part, $S$ the surface of the mobile part, $t_1$ (resp. $t_2$) the thickness of insulating layer 1 (resp. 2), $\varepsilon_1$ (resp. $\varepsilon_2$) the relative permittivity of layer 1 (resp. 2) and $\varepsilon$ the permittivity of the medium between the two electrodes.
Solving equation (1) yields the equilibrium position of the actuator mobile part provided that a solution exists.
This stable equilibrium actuation zone is defined by the whole actuator position $z$ in which a stable equilibrium occurs between the electrostatic force and the mechanical force.
The stable actuation zone can be modified to allow controlled motion in the entire gap and to overcome the *pull-in* effect after initial actuation gap so that, *pull-in* effect does not occurs, and the actuator motion is well controlled. The dimension condition which push-back the pull-in effect after the initial gap is defined as below [3]:

$$\left[ \frac{t_1}{\varepsilon_1} + \frac{t_2}{\varepsilon_2} \right] > \frac{2}{\varepsilon} \times g \quad (2)$$

If condition (2) is not observed, the equilibrium becomes unstable and the moving part collapses; This is the standard *pull-in* effect.
Indeed, if by construction, the previous condition is respected, then the displacement of the moving part can be controlled within the entire gap [4] [5].
For a conducting fluid, the voltage excitation has to be modulated in order to avoid electrode potential screening, as presented in [1]. Thus, the external drive voltage $V$, is assumed to be:

$$V = V_{rms} \sqrt{2} \cos(\omega t) \quad (3)$$

The induced electrostatic force has then a static component and a dynamic component at the pulsation *2ω*. In this case, quasi-static displacement of the actuator is achieved provided that mechanical response at pulsation *2ω* is negligible compared to the displacement at the resonance frequency. Under this assumption, the modelling we propose remains valid and can be used in case of AC driving voltage by simply replacing in Eq. (1) $V$ by the *rms* value of the AC voltage $V_{rms}$.
This enables to control the whole displacement of the actuator, which is very interesting for catching and seizing objects. Furthermore, making dynamical measurement by using the same device will be very helpful.

## 2.1. Dynamic study

In this section, we describe the analytical modeling of dynamic electrostatic actuation in liquid environment. The frequency response, the quality factor, the effective mass and the damping coefficient can be extracted from this model.
We consider a rectangular cantilever with a length, $L$, a thickness, $t$, and a width, $w$, electrostaticaly actuated at $g=2\mu m$ from the surface.
The differential equation for the motion $z(t)$ of the cantilever is:

$$m^* \frac{d^2 z(t)}{dt^2} = -kz(t) + \gamma \frac{dz(t)}{dt} + F_{elect} + F_{int} \quad (4)$$

Where, $m^*$, $\gamma$, and $k$ are, respectively, the effective mass, the damping coefficient and the cantilever stiffness. These parameters take into account the hydrodynamic effect [6] [7]:

$$m^* = \frac{\pi}{4} \rho \times w^2 \times L \times \Gamma_{rect-r}(\omega) + \rho_{lever} \times L \times w \times t \quad (5)$$

$$\gamma = \frac{\pi}{4} \times \rho \times \omega_{hyd} \times w^2 \times L \times \Gamma_{rect-i}(\omega) \quad (6)$$

$$Q = \frac{1}{\Gamma_{rect-i}} \times \left[ \frac{4}{\pi} \times \frac{\rho_{lever}}{\rho} \times \frac{t}{w} + \Gamma_{rect-r} \right] \quad (7)$$

Where $\rho$ is the density of the environment, $\omega_{hyd}$ the pulsation in the fluid, $\Gamma_{rect-r}(\omega)$ and $\Gamma_{rect-i}(\omega)$ are respectively the real and the imaginary part of the hydrodynamic function $\Gamma_{rect}(\omega) = \Gamma_{rect-r} + i \times \Gamma_{rect-i}$, defined as below [7]:

$$\Gamma_{rect-r} = 1.0553 + 3.7997 \times \frac{1}{w} \times \sqrt{\frac{2\eta}{\rho \omega_{hyd}}} \quad (8)$$

$$\Gamma_{rect-i} = 3.8018 \times \frac{1}{w} \sqrt{\frac{2\eta}{\rho \omega_{hyd}}} + 2.7364 \times \left[ \frac{1}{w} \sqrt{\frac{2\eta}{\rho \omega_{hyd}}} \right]^2 \quad (9)$$

Where, $\eta$ is the viscosity of the fluid.

$F_{elect}$ is the electrostatic force, and $F_{int}$ is the interaction force between the cantilever and the surface, also called water squeeze-out [8]:

$$F_{int} = \frac{\eta w^3}{g^3} \frac{dz(t)}{dt} \tag{10}$$

To solve the nonlinear differential equation above, we use the least action principle. The Lagrangian, $L(z, \dot{z}, t)$ writes:

$$L(z, \dot{z}, t) = T_k - U + W \tag{11}$$

Where $T_k$ is the kinetic energy, $U$ the potential energy and, $W$ the mechanical work. In our case, the Lagrangian becomes:

$$L(z, \dot{z}, t) = \frac{1}{2} m^* \dot{z}(t)^2 - \left[ \frac{1}{2} kz(t)^2 - E_{elect} \right] + z(t)\dot{z}(t) \left[ \gamma + \frac{\eta w^3}{g^3} \right] \tag{12}$$

Where $E_{elect}$ is the electrostatic energy and the underlined variable $\dot{z}(t)$ is calculated along the physical path, and thus is not varied in the calculations [9].

According to the assumption that the cantilever motion is harmonic at the working frequency, we use a trial function given by $z(t) = A \cos(\omega t + \phi)$. Here $A(t)$, and $\Phi(t)$ correspond respectively to the amplitude and the phase of the harmonic response. These two functions are assumed to have a slow time dependence compared to the period $T = 2\pi/\omega$. We then consider only the mean Lagrangian during one period, $<L>$, which appears as an effective Lagrangian for a large time scale compared to the period [10]:

$$\langle L(z, \dot{z}, t) \rangle = \frac{1}{T} \int_0^T L dt \tag{13}$$

Using the least action principle, we obtain the following set of equations:

$$\frac{\partial \langle L \rangle}{\partial A} = 0 \text{ and } \frac{\partial \langle L \rangle}{\partial \phi} = 0 \tag{14}$$

Solving these coupled equations gives relationships between the amplitude, the phase and the frequency. Thus theoretical frequency response, *A(f)* and *Φ(f)*, can be plotted as presented in Fig. 4. And then, the fundamental resonance frequency in liquid environment can be deduced:

$$f_{0-hyd-theo} = \frac{f_{0-vacuum-theo}}{\sqrt{1 + \frac{\pi}{4} \times \frac{\rho\ w}{\rho_{lever}} \Gamma_{rect-r}(\omega_{hyd})}} \tag{15}$$

Where $f_{0-vacuum-theo}$ is the fundamental resonance frequency in vacuum. In dynamic mode the fluids around the cantilever will increase the effective mass due to the hydrodynamic force and consequently the peak of the resonance frequency will decrease Fig. 4.

## 3. FABRICATION PROCESS

Three types of parallel plate actuators have been fabricated using the same technological process: a simple parallel plate device, a torsional actuator and finally a membrane clamped in its 4 corners, see Fig. 2.

Starting from a silicon wafer, a 350nm dry oxidation is performed and a 100nm $Si_xN_y$ (the exact stoichiometry has been optimized to reduce residual tensile stress) layer is deposited at 800°C. The purpose of this layer is to insulate devices from the substrate. A 350nm n-doped polycrystalline silicon layer is then deposited by Low Pressure Chemical Vapor Deposition (LPCVD), transferred by photolithography and $SF_6$ Reactive Ion Etching (RIE) to pattern buried electrodes. To insulate the electrodes during actuation, a 300nm $Si_xN_y$ layer is used, see Fig. 3-a. A 2µm sacrificial layer of Low Temperature Oxide (LTO) is deposited by LPCVD and anisotropically etched in a $CF_4$ and $CHF_3$ plasma, see Fig. 3-b.

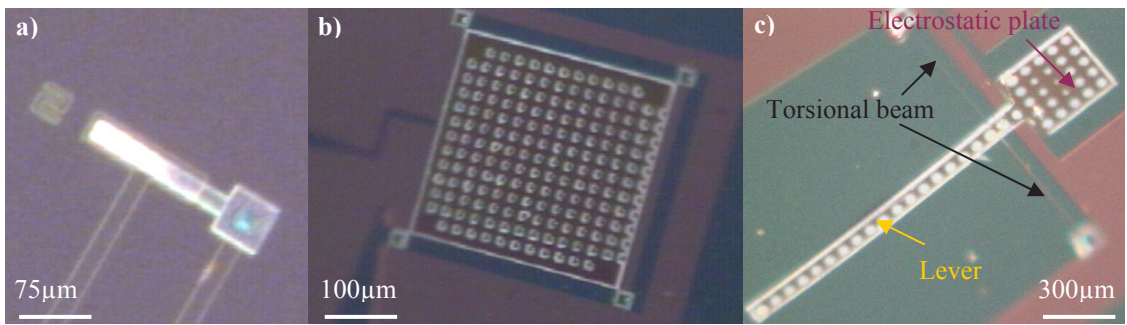

*Fig. 2 Optical microscope image during actuation in water. a) a 250µm long, 30µm width and 2µm thick beam actuator. b) a 378µm long, 243µm width and 2µm thick corner-clamped membrane. c) Torsional actuator with a 300µm square electrostatic plate for actuation and 490µm long and 5µm width torsional beam, deflection is observed on a 1100µm long and 90µm wide lever.*

Since actuators are operated in a liquid environment, encapsulation with $Si_xN_y$ is achieved to prevent any short-circuits or electrical leakages through the liquid medium. The breakdown field of $Si_xN_y$ deposited in our laboratory is $7 MV.cm^{-1}$, so that a 300nm thick $Si_xN_y$ layer is able to withstand a breakdown voltage of 200V.

The first level of encapsulation is made of a 300nm thick $Si_xN_y$ anisotropically etched in order to open the electric contacts, Fig. 3-c. The structural material is a 2µm n-doped polycrystalline silicon, which is anisotropically etched by $SF_6$ RIE, see Fig. 3-d. The second 300nm silicon nitride for the front side insulation is deposited and patterned resulting in the complete encapsulation of the device, see Fig. 3-e. In order to relax the mechanical stresses in the stacked layers, a thermal annealing step is performed at 1100°C during 3 hours. The last step consists in the release of the structure in a Hydrofluoric (HF) aqueous solution, see Fig. 3-f.

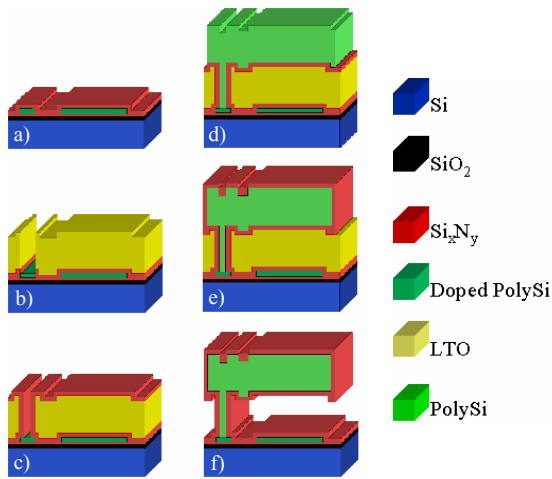

*Fig. 3 Schematic view of the process flow.*

Legend:
- Si
- $SiO_2$
- $Si_2N_y$
- Doped PolySi
- LTO
- PolySi

## 4. EXPERIMENTAL CHARACTERIZATION

### 4.1. Static study

Experiments have been performed in different liquids with various permittivity and conductivity such as air, IsoPropyl Alcohol (IPA) and tap water. The three types of actuator have been tested: torsional actuator, cantilever and membrane (see Fig. 2).

The parallel plate has been firstly tested in air since condition (2) is not respected and the pull-in effect occurs at 8V (Table 1). Rising the relative permittivity of the medium by actuating in IPA, (2) is not respected any longer and the pull-in effect appears at 4.1V (Table 1). The actuation voltage is reduced with a 50% factor compared to air medium. The last test occurs in tap water, where condition (2) is satisfied and motion of the mobile part is controlled in the entire gap range. Here the voltage needed to close the 2µm gap is around 6V (Table 1).

An interesting type of actuation in liquid environment is the torsional electrostatic actuation which could be used to make a micro-gripper. This device has been tested in air, in tap water and in IPA (Table 1). The $V_{pull-in}$ voltages are lower than the ones in the previous design, because of the higher surface of the actuation. As actuation occurs in a liquid environment, fluid displacement is involved.

To test the structures robustness, a membrane clamped in its 4 corners has been actuated in air, in tap water and in IPA (Table 1). We notice a good agreement between experimental and modeled values. This type of design presents a strong effective stiffness, which explains the high values of the pull-in voltages.

### 4.1. Dynamic study

Measurements have been performed in air and in water on the cantilever presented in Table 2. Our setup consists in a doppler laser and lock-in detection.

Theoretical (equation 14) and experimental fundamental resonance frequencies are presented in Fig. 4, we clearly see the frequency shift and the considerable decreasing of the quality factor.

| Devices | Actuation | Air | IsoPropyl Alcohol | Tap Water |
|---|---|---|---|---|
| | Relative permittivity | 1 | 21.3 | 80.1 |
| | $fc^d$ (kHz) | -- | 17 | 950 |
| Beam actuator | Theoretical [a] Voltage (V) | 7.6[*] | 3.9[*] | 6.3 |
| | Experimental [b] Voltage ($V_{rms}$) | 8[*] | 4.1[*] | 6 |
| Torsional actuator | Theoretical [a] Voltage (V) | 0.25[*] | 0.12[*] | 0.19 |
| | Experimental [b] Voltage ($V_{rms}$) | 0.3[*] | 0.16[*] | 0.23 |
| Membrane | Simulated [c] Voltage (V) | 44.68[*] | 16.37[*] | 33.4 |
| | Experimental [b] Voltage ($V_{rms}$) | 45[*] | 15.9[*] | 33.9 |

[a] Theoretical voltage to close the 2µm gap. [b] Experimental voltage to close the 2µm gap. [c] Simulated with CoventorWare™ software. [d] $f_c$ is the minimum signal frequency required to prevent electrode potential screening (calculated from[1]). [*]indicates that the pull-in effect occurs during actuation.

*Table 1 Theoretical and experimental actuation voltages required to close the gap for the cantilever, the torsional actuator and the membrane.*

| Parameter | Value |
|---|---|
| Cantilever length | $L = 250\mu m$ |
| Cantilever width | $w = 30\mu m$ |
| Cantilever thickness | $t = 2\mu m$ |
| Surface of the mobile part | $S = 7500\mu m^2$ |
| Gap | $g = 2\mu m$ |
| Thickness of insulating $Si_xN_y$ | $t_1 = t_2 = 300nm$ |
| Relative permittivity of water | $\varepsilon = 80.1$ |
| Relative permittivity of $Si_xN_y$ [a] | $\varepsilon_1 = 8$ |
| Viscosity of water | $\eta = 8.59\times10^{-4} kg.m^{-1}.s^{-1}$ |
| Density of PolySi [a] | $\rho_{lever} = 2330 \, kg.m^{-3}$ |
| Vacuum dielectric permittivity | $\varepsilon_0 = 8.85\times10^{-12} F.m^{-1}$ |

[a] Polycristalline silicon.

*Table 2 : Parameters of the cantilever.*

Table 3 presents theoretical and experimental results in vacuum, air and tap water. We find a concordance on the frequency of around 8% in water and 1.5% in the case of actuation in air, which denotes a good agreement between the analytical modeling and the tested values.

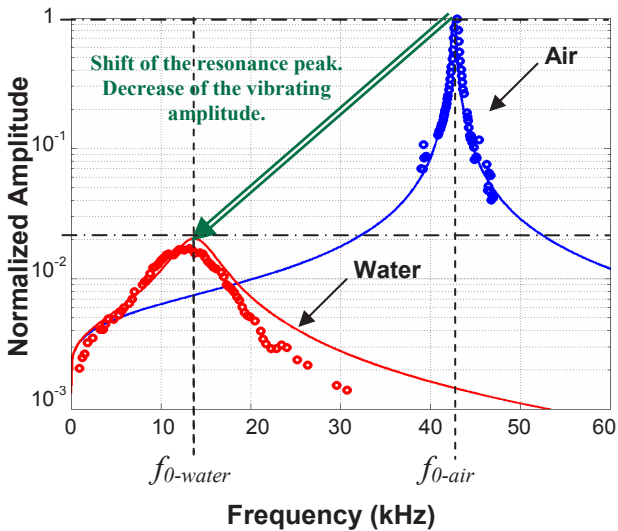

*Fig. 4 Fundamental frequency response in air and in water of the cantilever (dimensions in Table 2) showing a shift of resonance peak of 67%. Vibrations amplitudes of response are normalized so that the amplitudes in air is equal to 1. Experimental data are plotted with circle.*

As damping effect are important in fluids, see Table 3, the theoretical transient mode is also presented in Fig. 7, this shows that we have to take into account the stabilization time before making measurement.

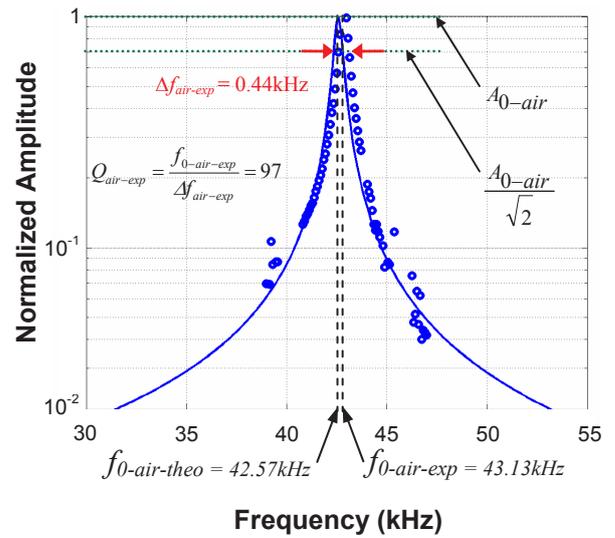

*Fig. 5 Comparison between the experimental (circle) and the theoretical (continuous line) resonance peak of the beam in air. The theoretical values predicts a $Q_{air-theo}$=98.31 at $f_{0-air-theo}$=42.57kHz.*

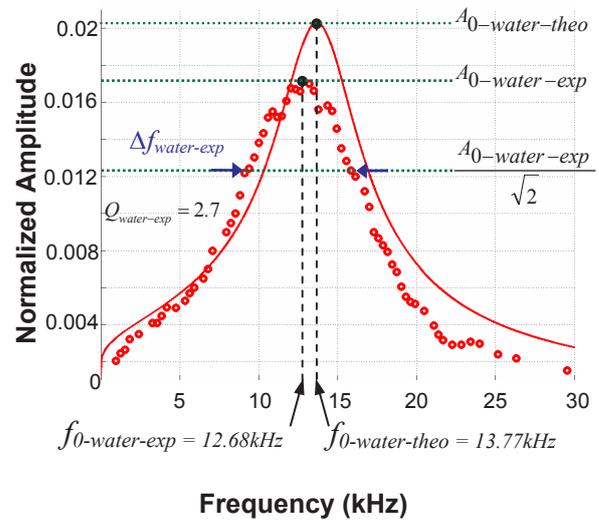

*Fig. 6 Comparison between the experimental (circle) and the theoretical (continuous line) resonance peak of the beam in water. The theoretical values predicts a $Q_{water-theo}$=3.19 at $f_{0-water-theo}$=13.77kHz.*

| Environment | Fundamental resonance frequency (kHz) | Quality factor[a] | Effective mass (kg) | Viscous damping (kg.rad.s⁻¹) |
|---|---|---|---|---|
| Vacuum (theoretical) | 42.88 | ∞ | $3.49 \times 10^{-11}$ | 0 |
| Air (theoretical) | 42.57 | 98.31 | $3.5456 \times 10^{-11}$ | $9.69 \times 10^{-8}$ |
| Air (tested) | 43.13 | 97 | $3.5454 \times 10^{-11}$ | $9.72 \times 10^{-8}$ |
| Water (theoretical) | 13.77 | 3.19 | $3.09 \times 10^{-10}$ | $1.06 \times 10^{-5}$ |
| Water (tested) | 12.68 | 2.7 | $3.24 \times 10^{-10}$ | $9.19 \times 10^{-6}$ |

[a] For theoretical values, we assume that the damping depends only on hydrodynamic effect.

*Table 3 : Theoretical and experimental resonance frequency, quality factor, damping and effective mass for a 250µm long, 30µm width and 2µm thick cantilever encapsulated with 300nm of $Si_xN_y$.*

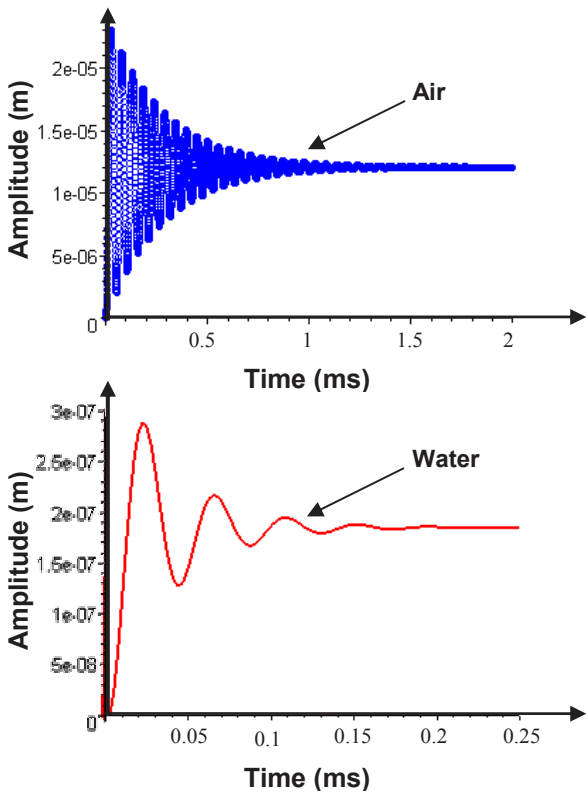

*Fig. 7 Comparison of the amplitude variation with time in air and in water.*

## 5. CONCLUSION

In this paper, we have presented a novel and optimized fabrication technique of parallel plate actuators with a complete $Si_xN_y$ encapsulation, allowing static and dynamic electrostatic actuation in a liquid environment. Static analytical calculations show that the pull-in effect can be shifted far beyond one-third of the gap and even suppressed. Experiments have been performed on fabricated devices and actuation has been successfully observed on cantilevers, torsional actuators and membranes. The voltages required to close the actuators gap have been measured in air, IPA and Tap water in good agreement with the theoretical values.

Furthermore, the dynamic measurements in air and in water are in good agreement with the hydrodynamic analytical modeling of actuation. In dynamic mode, this analytical modeling gives guidelines to design cantilever actuating in liquid environment. In particular, this model enables an optimization of the frequency and the quality factor for AFM imaging in liquid solutions in particularly for submicroscopic object like cellular components.